\documentclass[12pt]{article}
\usepackage[utf8]{inputenc}
\usepackage{amsmath,amsfonts,amssymb}
\usepackage[T2A]{fontenc}
\usepackage[width=18cm,height=24cm]{geometry}
\usepackage{xcolor}
\usepackage{hyperref}
\usepackage{cite}
\usepackage{bm}
\numberwithin{equation}{section}
\usepackage[titletoc,title]{appendix}

\def\mathclap#1{\text{\hbox to 0pt{\hss$\mathsurround=0pt#1$\hss}}}

\newcommand{\D}{\mathbb{D}}
\newcommand{\dd}{\mathrm{d}}
\newcommand{\E}{\mathcal{E}}
\newcommand{\al}{{\alpha}}
\newcommand{\be}{{\beta}}
\newcommand{\g}{{\gamma}}
\newcommand{\ad}{{\dot{\alpha}}}
\newcommand{\bd}{{\dot{\beta}}}
\newcommand{\gd}{{\dot{\gamma}}}

\newcommand{\deld}{{\dot{\delta}}}
\newcommand{\y}{{\bar{y}}}
\newcommand{\dif}{{\bar{\partial}}}

\newcommand{\rhs}{\textit{r.h.s.}}

\def\mathclap#1{\text{\hbox to 0pt{\hss$\mathsurround=0pt#1$\hss}}}
\newcommand{\fint}[1]{\int\limits_\mathclap{{\hphantom{#1} #1}}}

\numberwithin{equation}{section}

\begin{document}
\begin{titlepage}

\begin{flushright}
\vspace{1mm}
 FIAN/TD/8--2026\\
\vspace{-1mm}
\end{flushright}\vspace{5cm}

\begin{center}
{\LARGE\bf Topological Fields in $4d$ Higher Spin Theory}

\vspace{2 cm}

{\large\bf P.T.~Kirakosiants}\\

\vspace{0.5 cm}
I.E. Tamm Department of Theoretical Physics, Lebedev Physical Institute, \\ Leninsky prospect 53, 119991, Moscow, Russia

\end{center}
\vspace{1 cm}

\begin{abstract}

The equations for topological fields in the $4d$ higher spin theory are considered. It is shown that these fields contain a finite number of degrees of freedom that justifies their naming. The issue of construction of gauge invariant functionals is addressed, and a gauge-invariant cubic action is constructed for the interacting physical and topological higher spin fields.

\end{abstract}

\end{titlepage}

\newpage

\tableofcontents
\newpage

\section{Introduction}
\label{introduction}

Higher Spin (HS) gauge theory is of significant interest in theoretical physics \cite{Bekaert:2004qos,Bekaert:2010hw,Ponomarev:2022vjb}. Its aim is to describe gauge (massless) fields with arbitrary integer and half-integer spins. While HS fields have been described at the free-field level by Fronsdal and Fang \cite{Fronsdal:1978rb,Fang:1978wz}, questions regarding their interactions remain actively discussed (see \cite{Didenko:2026nag} and references therein). An interesting area of research is the investigation of a possible connection between the theories of massless HS fields and string theory \cite{Sagnotti:2011jdy,Gaberdiel:2015wpo,Vasiliev:2018zer,Tarusov:2025sre,Didenko:2025xca,Tarusov:2026ich}. In this paper, we focus on a system of generating equations \cite{Vasiliev:1992av} that enables the derivation of interaction vertices for HS fields at all orders. It is an important feature of this system that the fields can depend on the so-called Klein operators in various ways, which results in their decomposition into physical and topological fields, for instance, in the case of fluctuations on the $AdS_4$-background. The conventional approach involves a consistent reduction, which leaves only physical HS fields (however, such a reduction is not possible in three dimensions, and related questions are currently of interest, see e.g. \cite{Vasiliev:2025erl}). We relax such a reduction and consider how topological fields contribute to the theory in the lowest orders. In addition to its independent significance, this consideration is motivated from the perspective that similar (though more complex) issues arise in the context of generalizations involving Coxeter groups, which is believed to be related to string theory \cite{Vasiliev:2018zer,Tarusov:2025sre,Tarusov:2026ich}.

It will be demonstrated that free equations for topological fields without a \rhs\,  possess solutions that can be removed by gauge transformations.
This check is necessary since the gauge fields are valued in the infinite-dimensional
(twisted adjoint) representation of the $AdS_4$ algebra, that in principle may obscure
formal gauge fixing manipulations.

When Weyl tensor-like terms appear on the \rhs\, as a consequence of the nonlinear system, these solutions acquire a non-trivial form, although they still do not possess field degrees of freedom and therefore describe topological fields. A gauge invariant cubic action for the interaction of physical and topological HS fields will be constructed along with the charges resulting from this action. Note that, as discussed in \cite{Didenko:2015pjo}, topological fields can be interpreted as coupling constants (moduli) of  the HS theory. From that perspective, their
contribution and symmetry may be important in various respects.

Note that an analogous field system was considered in \cite{Vasiliev:1987hv} within the Dirac approach to constrained Hamiltonian systems. It was shown therein that the fields in question carry no field degrees of freedom thus being topological. In this work, we demonstrate this in terms of the unfolded equations of motion, that simplifies the analysis.

The rest of the paper is organized as follows. The equations under consideration and their origin from the nonlinear system of equations for HS fields are discussed in Section \ref{sec2}. Homogeneous linear equations on topological one-form fields are investigated in Section \ref{sec3}. The linear equations with nonzero \rhs\, are considered in Section \ref{sec4}. The action for interacting HS and topological fields with conserved charges are constructed in Section \ref{sec5}. In addition, there are two appendices that provide technical details.

\section{Generating system}
\label{sec2}

The fruitful approach to describe HS fields is the following \cite{Vasiliev:1988sa}. Physical gauge fields of integer spins $s \geq 1$ are described by space-time one-forms $\omega_{\al(s-1),\ad(s-1)}$  ($\al, \ad \in \{ 1, 2 \}$)\footnote[1]{Following \cite{Vasiliev:1987hv}, we use the short-hand notation for symmetrized two-component spinor indices:
\begin{equation}
    A_{\al(m)} \equiv A_{\underbrace{\al \dots \al}_{m}} = \frac{1}{m!}\sum_{\sigma\in S_m} A_{\al_{\sigma(1)}\dots \al_{\sigma(m)}}, \quad S_m \text{ is the symmetric group,}
\label{Xeqn1-2.1}
\end{equation}
and we assume that indices denoted by the same letter are symmetrized independently with respect to their position (upper or lower). After this symmetrization, the maximum possible number of upper and lower indices denoted by the same letter must be contracted.}.  Gauge fields of half-integer spins $s \geq \frac{3}{2}$ are described by space-time one-forms $\omega_{\al(n),\ad(n-1)}$ and $\omega_{\al(n-1),\ad(n)}$ with $2n = 2s-1$. Scalar matter fields are zero-forms $C$ (without spinor indices), and spin-$\frac{1}{2}$ fields are described by zero-forms $C_{\al}$ and $\bar{C}_{\ad}$. These fields are packed into
\begin{equation}
    \omega(y,\y|x) = \sum_{n, m = 0}^\infty \frac{1}{n!m!} \omega_{\al_1 \dots \al_n,\ad_1 \dots \ad_m} y^{\al_1} \dots y^{\al_n} \y^{\ad_1} \dots \y^{\ad_m}
    \label{omega_expansion}
\end{equation}
and
\begin{equation}
    C(y,\y|x) = \sum_{n, m = 0}^\infty \frac{1}{n!m!} C_{\al_1 \dots \al_n,\ad_1 \dots \ad_m} y^{\al_1} \dots y^{\al_n} \y^{\ad_1} \dots \y^{\ad_m},
    \label{C_expansion}
\end{equation}
where auxiliary spinor variables $Y^A = (y^\al, \y^\ad)$ with the conjugating rule $(y^\al)^{\dagger} = \y^\ad$ are introduced. Components $\omega_{\al_1 \dots \al_n,\ad_1 \dots \ad_m}$, $C_{\al_1 \dots \al_n,\ad_1 \dots \ad_m}$, which were not defined above as physical fields, are expressed in terms of derivatives of physical fields due to the unfolded HS equations \cite{Vasiliev:1988sa} (Central On-shell theorem):
\begin{align}
    &D^L \omega + h^{\al\ad}\big( y_\al \bar{\partial}_\ad + \y_\ad \partial_\al \big) \omega = \frac{i\eta}{2}\bar{H}_{\dot{\alpha}\dot{\beta}}\bar{\partial}^{\dot{\alpha}}\bar{\partial}^{\dot{\beta}}C(0,\bar{y}) + \frac{i\bar{\eta}}{2}H_{{\alpha}{\beta}}\partial^\al \partial^\be C(y, 0),
    \label{TheFirstOn-shell:omega}
\\
    &D^L C - ih^{\al \ad}(y_\al \y_\ad - \partial_\al \dif_\ad)C = 0.
    \label{TheFirstOn-shell:C}
\end{align}
Here,  $\eta$ is a complex parameter, $h^{\al \ad}$ is a part of $AdS$-connection:
\begin{equation}
    \omega_{AdS} = -\frac{i}{4}\big( \omega_{\al\be} y^\al y^\be + \bar{\omega}_{\ad\bd} \y^\ad \y^\bd + 2 h_{\al\ad}y^\al \y^\ad \big),
    \label{AdS-connection}
\end{equation}
and the Lorenz covariant derivative acts as
\begin{equation}
    D^L = \dd_x + \omega_{\al\be}y^\al \partial^\be + \bar{\omega}_{\ad\bd} \y^\ad \dif^\bd,
\label{Xeqn7-2.7}
\end{equation}
\begin{equation*}
    \dd_x = \dd x^n \frac{\partial}{\partial x^n}, \quad \partial_\al = \frac{\partial}{\partial y^\al}, \quad \dif_\ad = \frac{\partial}{\partial \y^\ad}.
\end{equation*}
The two forms $H_{\al\be}, \bar{H}_{\ad\bd}$ are
\begin{equation}
    H_{\al\be} = h_{\al\ad}{h_\be}^\ad, \quad \bar{H}_{\ad\bd} = h_{\al\ad}{h^\al}_\bd,
\label{Xeqn8-2.8}
\end{equation}
and we use a symplectic form on spinor space:
\begin{equation}
    \epsilon_{AB}=(\epsilon_{\alpha \beta}, \epsilon_{\dot{\alpha} \dot{\beta}}): \quad  \epsilon_{\alpha \beta}= - \epsilon_{\beta \alpha}, \quad \epsilon _{\dot{\alpha} \dot{\beta}} = -\epsilon_{\dot{\beta} \dot{\alpha}}, \quad \epsilon_{12} = 1
\end{equation}
with the following convention on raising and lowering indices:
\begin{equation}
    A_{\alpha} = A^{\beta}\epsilon_{\beta \alpha}, \quad A^{\alpha} = \epsilon^{\alpha \beta}A_{\beta}, \quad A_{\dot{\alpha}} = A^{{\dot\beta}}\epsilon_{{\dot\beta} {\dot\alpha}}, \quad A^{\dot{\alpha}} = \epsilon^{\dot{\alpha} {\dot\beta}}A_{{\dot\beta}}.
\label{Xeqn10-2.10}
\end{equation}

{\eqref{TheFirstOn-shell:omega}} and \eqref{TheFirstOn-shell:C} contain dynamic equations on free HS fields along with expressions for the other components of \eqref{omega_expansion}, \eqref{C_expansion} that are expressed via derivatives of the
former.

The problem to be solved within the generating system is the construction of nontrivial interacting vertices in the \rhs\, of \eqref{TheFirstOn-shell:omega} and \eqref{TheFirstOn-shell:C}. It is solved as follows \cite{Vasiliev:1992av}. The space of spinor variables is extended by $Z^A = (z_\al, \bar{z}_\ad)$ with the differential $\dd Z^A = \theta^A = (\theta^\al, \bar{\theta}^\ad)$. There is the associative star product:
\begin{equation}
    f(Z, Y) * g(Z, Y) = \int \frac{\dd^4 U \dd^4 V}{(2\pi)^4} f(Z+U, Y+U) g(Z-V, Y+V) e^{i U_A V^B}.
    \label{star}
\end{equation}
The outer Klein operators $K = (k, \bar{k})$ are introduced as follows:
\begin{align}
    \{k, q_{\alpha}\} &= 0, \quad [k, \bar{q}_{\dot{\alpha}}] = 0, \quad \{\bar{k}, \bar{q}_{\ad}\} = 0, \quad [\bar{k}, q_\al] = 0, \notag\\
    k^2 &= \bar{k}^2 = 1, \quad [k, \bar{k}]=0, \quad k^{\dagger} = \bar{k},
\end{align}
where $Q_A = (q_\al, \bar{q}_\ad)$ stands for $Y_A, Z_A$ or $\theta_A$.

The generating system is
\begin{subequations}
\label{base_all}
\begin{align}
    &\dd_x W + W * W = 0, \label{base_eq1} \\
    &\dd_x B + W* B - B * W = 0, \label{base_eq2} \\
    &\dd_x S + W * S + S * W= 0, \label{base_eq3} \\
    &S * B - B * S = 0, \label{base_eq4} \\
    &S * S = i(\theta^{\alpha} \theta_{\alpha} + \bar{\theta}^{\dot{\alpha}} \bar{\theta}_{\dot{\alpha}} + B * (\eta \gamma + \bar{\eta} \bar{\gamma})),\label{base_eq5}
\end{align}
\end{subequations}
where $\eta$ is a complex parameter,
\begin{equation}
    \gamma = \theta^\al\theta_\al e^{i z_\al y^\al}k, \quad \bar{\gamma} =\bar{\theta}^{\dot{\alpha}} \bar{\theta}_{\dot{\alpha}} e^{i \bar{z}_\ad \y^\ad} \bar{k }
\label{Xeqn12-2.14}
\end{equation}
are central elements (i.e. they commute with each function), and a set of fields is specified as follows: $W = W_n(Z,Y,K|x) \dd x^n$ and $S = S_A(Z,Y,K|x) \theta^A$ are one-forms, $B = B(Z,Y,K|x)$ is a zero-form.

The system \eqref{base_all} reproduces perturbatively interacting HS vertices. The vacuum solution is
\begin{equation}
    W_0 = \omega_{AdS}, \quad B_0 = 0, \quad S_0 = \theta^{\alpha} z_{\alpha} + \bar{\theta}^{\dot{\alpha}}\bar{z}_{\dot{\alpha}},
    \label{vac}
\end{equation}
where $\omega_{AdS}$ is an $AdS-$connection \eqref{AdS-connection} that satisfies the following flatness condition:
\begin{equation}
    \dd_x \omega_{AdS} + \omega_{AdS}*\omega_{AdS} = 0.
    \label{flatness}
\end{equation}

The following property of $S_0$:
\begin{equation}
    S_0 * f(Z) - f(Z) * S_0  = -2i \dd_Z f(Z), \quad \dd_Z = \theta^A \frac{\partial}{\partial Z^A},
    \label{d_z-prop}
\end{equation}
reduces the analysis of the system \eqref{base_eq1}--\eqref{base_eq5} to solving equations of the form
\begin{equation}
    \dd_Z f(Z, Y, \theta) = g(Z, Y, \theta).
    \label{eq_dd}
\end{equation}
In particular, from \eqref{base_eq4} it follows that the first correction $B_1$ to the field $B$ is $Z-$independent: $B_1 = C(Y,K|x)$. \eqref{base_eq2} implies
\begin{equation}
    \dd_x C + \omega_{AdS} * C - C*\omega_{AdS} = 0.
\label{Xeqn17-2.19}
\end{equation}

There are two possible types of dependence of $C$ on $K$. If it is odd, i.e. $C(Y,-K|x) = -C(Y,K|x)$, then $C$ satisfies the equation {\eqref{TheFirstOn-shell:C}} and contains HS fields and their derivatives. If $C$ is even in $K$, i.e. $C(Y,-K|x) = C(Y,K|x)$, it satisfies
\begin{equation}
    D^L C + h^{\al\ad}\big( y_\al \bar{\partial}_\ad + \y_\ad \partial_\al \big) C = 0.
    \label{eq:C}
\end{equation}

The equation {\eqref{base_eq3}} on the first correction of $W_1$ is
\begin{equation}
    2i\dd_Z W_1 = \dd_x S_1 + \omega_{AdS} *S_1 + S_1*\omega_{AdS}.
\label{Xeqn19-2.21}
\end{equation}
Note that a general solution to this equation contains an arbitrary $Z-$independent function $\omega$, which encodes the HS physical or topological fields. Then the substitution of $W_1$ into \eqref{base_eq1} yields the equation on the one-form fields. We face two options for the field $\omega$, depending on whether it is even or odd in $K$. Leaving details for Appendix \ref{App1}, we assert that the first option leads to \eqref{TheFirstOn-shell:omega} while the second one leads to
\begin{equation}
    D^L \omega - ih^{\al \ad}(y_\al \y_\ad - \partial_\al \dif_\ad)\omega = \frac{i\eta}{2}\bar{H}_{\dot{\alpha}\dot{\beta}}\bar{y}^{\dot{\alpha}}\bar{y}^{\dot{\beta}}C(0,\bar{y},K|x)k + \frac{i\bar{\eta}}{2}H_{{\alpha}{\beta}}y^\al y^\be C(y, 0,K|x)\bar{k},
    \label{eq:omega}
\end{equation}
where $C$ on the \rhs\, is even in $K$. It is not hard to see that \eqref{eq:omega} is invariant under the following gauge transformation:
\begin{equation}
    \delta \omega = D^L \epsilon - ih^{\al \ad}(y_\al \y_\ad - \partial_\al \dif_\ad)\epsilon
    \label{gauge:omega}
\end{equation}
with $\epsilon$ odd in $K$.

In this paper, we analyze the equations {\eqref{eq:C}} and \eqref{eq:omega}.

\section{Homogeneous equation}
\label{sec3}

Consider {\eqref{eq:omega}} with zero on the \rhs,
\begin{equation}
    D^L \omega - ih^{\al \ad}(y_\al \y_\ad - \partial_\al \dif_\ad)\omega = 0\,.
    \label{eq:omega_homog}
\end{equation}

The {\eqref{eq:omega_homog}} allows one to express all components $\omega_{\al(n),\ad(m)}$ in terms of $\omega_{\al(n)}$ and $\omega_{\ad(m)}$. Consider the equation on $\omega_{\al(n)}$ with $n > 0$ ($\omega_{\ad(n)}$ can be considered analogously):
\begin{equation}
    D^L \omega_{\al(n)} + i h^{\al\ad} \omega_{\al(n+1),\ad} = 0
\label{Xeqn22-3.2}
\end{equation}
with the gauge transformation
\begin{equation}
    \delta \omega_{\al(n)} = D^L \epsilon_{\al(n)} + i h^{\al\ad} \epsilon_{\al(n+1),\ad}.
    \label{gauge:spin=n}
\end{equation}
$\omega_{\al(n)}$ can be decomposed in the frame one-forms $h_{\al\ad}$ as
\begin{equation}
    \omega_{\al(n)} = h^{\al\ad}f_{\al(n+1),\ad} + {h_{\al}}^\ad f_{\al(n-1),\ad}.
    \label{spin=1:one-form-decomp}
\end{equation}

Let us contract the one-form index $m$ of $\delta \omega_{\al(n)} = \dd x^m \delta \omega_{m|\al(n)}$ in \eqref{gauge:spin=n} with the inverse frame field $h^m_{\al\ad}$ (defined in such a way that $h^m_{\al\ad} h_m^{\be\bd} = 2 \delta^\be_\al \delta^\bd_\ad$ and $h^m_{\al\ad} h_n^{\al\ad} = 2\delta^m_n$),
\begin{equation}
    \delta f_{\al(n-1),\ad} = -\frac{n}{2(n+1)}{h^{m\al}}_{\ad} D^L_m \epsilon_{\al(n)}
    \label{delta_f}\,.
\end{equation}
It may look like this gauge transformation allows one to eliminate all degrees of freedom in $f_{\al(n-1),\ad}$. But this is not the case, as is evident from the following consequence of \eqref{delta_f} for $n \geq 2$:
\begin{equation}
    h^{l\al\ad} D^L_l \delta f_{\al(n-1),\ad} = -\frac{n}{2(n+1)}h^{l\al\ad} D^L_l {h^{m\al}}_{\ad} D^L_m \epsilon_{\al(n)} = 0
    \label{equality_for_gauge}
\end{equation}
(see Appendix {\ref{App2} with the proof), from where it follows that $f_{\al(n-1),\ad}$ contains degrees of freedom that cannot be gauged away by \eqref{delta_f}. But these degrees are not allowed as a consequence of \eqref{eq:omega_homog}. In fact, from the decomposition
\begin{equation}
    D^L f_{\al(n-1),\ad} = h_{\al\ad} F_{\al(n-2)} + {h_\al}^\ad F_{\al(n-2),\ad\ad} + {h^\al}_\ad F_{\al(n)} + h^{\al\ad}F_{\al(n),\ad\ad},
    \label{Df_decomp}
\end{equation}
we can see that
\begin{equation}
    h^{l\al\ad} D^L_l f_{\al(n-1),\ad}  = \frac{4n}{n-1} F_{\al(n-2)},
    \label{hDf}
\end{equation}
and from \eqref{eq:omega_homog}, it follows that $F_{\al(n-2)}=0$. That is why \eqref{equality_for_gauge} does not preclude the possibility to gauge away $f_{\al(n-1),\ad}$.

The component $f_{\al(n+1),\ad}$ in \eqref{spin=1:one-form-decomp} can be gauged away by the appropriate choice of $\epsilon_{\al(n+1),\ad}$ in \eqref{gauge:spin=n}.

After these gauge-fixing steps we are left with the equation
\begin{equation}
    i h^{\al\ad} \omega_{\al(n+1),\ad} = 0.
    \label{eq:omega(n+1)1}
\end{equation}

We assert that this equation entails triviality of $\omega_{\al(n+1),\ad}$. To prove this statement, consider a one-form $\omega_{\al(n),\ad(m)}$ with $n > 0$, $m > 0$, $mn \neq 1$ and suppose that $\omega_{\al(n-k),\ad(m-k)} = 0$ for all possible $k > 0$. Then $\omega_{\al(n),\ad(m)}$ satisfies the equation
\begin{equation}
    i h^{\al\ad} \omega_{\al(n),\ad(m)} = 0.
    \label{eq:omega(nm)}
\end{equation}
The gauge transformation of $\omega_{\al(n),\ad(m)}$ is
\begin{equation}
    \delta \omega_{\al(n),\ad(m)} = D^L \epsilon_{\al(n),\ad(m)} - i nm h_{\al\ad} \epsilon_{\al(n-1),\ad(m-1)} + ih^{\al\ad} \epsilon_{\al(n+1),\ad(m+1)}.
    \label{gauge:omega(nm)}
\end{equation}

Decomposing $\omega_{\al(n),\ad(m)}$ as
\begin{align}
    \omega_{\al(n),\ad(m)}& = h^{\al\ad}f_{\al(n+1),\ad(m+1)} + {h_{\al}}^\ad f_{\al(n-1),\ad(m+1)} + {h^{\al}}_\ad f_{\al(n+1),\ad(m-1)} + h_{\al\ad}f_{\al(n-1),\ad(m-1)},
    \label{n,m:one-form-decomp}
\end{align}
we obtain
\begin{multline}
    i h^{\al\ad} \omega_{\al(n),\ad(m)} = -i \bar{H}^{\bd\bd} \big( a_1(n) f_{\al(n-1),\bd\bd\ad(m-1)} + a_2(n, m) \varepsilon_{\bd \ad} f_{\al(n-1),\bd\ad(m-2)} \big) -\\ -i H^{\be\be} \big( a_1(m) f_{\be\be\al(n-1),\ad(m-1)} + a_2(m, n) \varepsilon_{\be \al} f_{\be\al(n-2),\ad(m-1)} \big) = 0,
    \label{eq:2-forms-comp}
\end{multline}
where
\begin{equation}
    a_1(n) = \frac{n+1}{2n}, \quad a_2(n, m) = \frac{(n+1)(m-1)}{2nm}.
\label{Xeqn32-3.14}
\end{equation}

Comparing the coefficients in front of basis two-forms $H^{\be\be}$ and $\bar{H}^{\bd\bd}$, one finds
\begin{equation}
     a_1(n) f_{\al(n-1),\bd\bd\ad(m-1)} + a_2(n, m) \varepsilon_{\bd \ad} f_{\al(n-1),\bd\ad(m-2)} = 0,
     \label{eq:basis_coef_2holom}
\end{equation}
\begin{equation}
    a_1(m) f_{\be\be\al(n-1),\ad(m-1)} + a_2(m, n) \varepsilon_{\be \al} f_{\be\al(n-2),\ad(m-1)} = 0.
    \label{eq:basis_coef_2anti-holom}
\end{equation}
If $m \neq 1$, contraction of \eqref{eq:basis_coef_2holom} with $\varepsilon^{\bd\ad}$ leads to $f_{\al(n-1),\ad(m-1)} = 0$, as the first term in \eqref{eq:basis_coef_2holom} is symmetric with respect to the dotted indices. Therefore, $f_{\al(n-1),\ad(m+1)} = 0$ and from \eqref{eq:basis_coef_2anti-holom} it follows that $f_{\al(n+1),\ad(m-1)} = 0$ as well. If $m = 1$, then $n \neq 1$ (recall that $nm > 1$), and we follow similar steps starting with \eqref{eq:basis_coef_2anti-holom} to arrive at the same conclusion.

Thus we conclude that only the first term in \eqref{n,m:one-form-decomp} is non-trivial. But it can be eliminated by the gauge transformation \eqref{gauge:omega(nm)} with an appropriate $\epsilon_{\al(n+1),\ad(m+1)}$. Therefore, $\omega_{\al(n),\ad(m)}$ is pure gauge for $n > 0, m > 0, nm > 1$.

It remains to consider the equation for $\omega$ with $n = m = 0$. In that case $\omega = h^{\al\ad}f_{\al\ad}$ and hence
\begin{equation}
    D^L \omega + ih^{\al \ad}\omega_{\al\ad} = 0
    \label{eq:spin0}
\end{equation}
with the gauge transformation
\begin{equation}
    \delta \omega = D^L \epsilon + ih^{\al \ad}\epsilon_{\al\ad}.
\label{Xeqn36-3.18}
\end{equation}
Choosing the gauge $ \epsilon^0_{\al\ad} = i f_{\al\ad}$, we are left with $\omega = 0$ and the following conditions on the remaining gauge parameters:
\begin{equation}
    D^L \epsilon + ih^{\al \ad}\epsilon_{\al\ad} = 0,
\label{Xeqn37-3.19}
\end{equation}
from which it follows that
\begin{equation}
    \epsilon_{\al\ad} = \frac{i}{2} h^n_{\al\ad}D^L_n \epsilon.
    \label{epsilon_alad}
\end{equation}

Thus, \eqref{eq:spin0} reduces to
\begin{equation}
    h^{\al \ad}\omega_{\al\ad} = 0.
    \label{eq:spin0_rem}
\end{equation}
The following decomposition holds:
\begin{equation}
    \omega_{\al\ad} = h^{\al\ad}f_{\al(2),\ad(2)}+{h_\al}^\ad f_{\ad(2)}+{h^\al}_\ad f_{\al(2)} + h_{\al\ad}f,
    \label{decomp:omega(1,1)}
\end{equation}
and the gauge transformation for $\omega_{\al\ad}$ is
\begin{equation}
    \delta \omega_{\al\ad}  =D^L \epsilon_{\al\ad} - ih_{\al\ad}\epsilon + i h^{\al\ad}\epsilon_{\al(2),\ad(2)}.
    \label{gauge:omega(1,1)}
\end{equation}

Substitution of \eqref{decomp:omega(1,1)} into \eqref{eq:spin0_rem} leads to $f_{\al(2)} = f_{\ad(2)} = 0$. Contraction of \eqref{gauge:omega(1,1)} with the inverse frame field ${h^n}^{\al\ad}$ taking into account \eqref{epsilon_alad} entails
\begin{equation}
    \delta f  = \frac{i}{16} h^{n\al\ad}D^L_{n}{h^m}_{\al\ad}D^L_m \epsilon - i \epsilon.
\label{Xeqn42-3.24}
\end{equation}

An appropriate choice of $\epsilon$ allows one to eliminate $f$ (locally). The leftover component $f_{\al(2),\ad(2)}$ in \eqref{decomp:omega(1,1)} can be eliminated by the corresponding $\epsilon_{\al(2),\ad(2)}$ in \eqref{gauge:omega(1,1)}.

The inductive analysis above shows that the system \eqref{eq:omega_homog} leads to $\omega = 0$ modulo a gauge transformation.

\section{Inhomogeneous equation}
\label{sec4}

Now consider \eqref{eq:omega} with non-zero \rhs. In that case the gauge transformation \eqref{delta_f} does not allow to eliminate $f_{\al(n-1),\ad}$ for $n \geq 2$, since $h^{l\al\ad} D^L_l f_{\al(n-1),\ad} \neq 0$. Actually, recall the decomposition \eqref{Df_decomp}. The first term in \eqref{Df_decomp} is determined by the equation {\eqref{eq:omega}}:
\begin{equation}
    F_{\al(n-2)} = \frac{i\bar{\eta}}{2} C_{\al(n-2)}.
\label{Xeqn43-4.1}
\end{equation}
From \eqref{hDf} it follows that
\begin{equation}
   h^{l\al\ad} D^L_l f_{\al(n-1),\ad}  = \frac{2n}{n-1}i\bar{\eta} C_{\al(n-2)}.
   \label{Df=C}
\end{equation}

A general solution to
\eqref{Df=C} is $f^{(0)}_{\al(n-1),\ad} + f^{(1)}_{\al(n-1),\ad}$, where $f^{(0)}_{\al(n-1),\ad}$ is the solution to the homogeneous equation (with $C_{\al(n-2)} = 0$) and $f^{(1)}_{\al(n-1),\ad}$ is the particular solution to the inhomogeneous equation. From the previous section, we know that $f^{(0)}_{\al(n-1),\ad}$ can be gauged away. Therefore, only $f^{(1)}_{\al(n-1),\ad}$ remains nontrivial after gauge fixing. It will be demonstrated below that $C_{\al(n-2)}$, which generates $f^{(1)}_{\al(n-1),\ad}$, has a finite number of degrees of freedom. Thus, the system does not carry field degrees of freedom.

This can be understood as follows. Let $M$ denote the space of degrees of freedom in a homogeneous equation. There is a group of gauge symmetry transformations that act transitively on this space, meaning that for each point in $M$, there exists an appropriate gauge transformation that brings it to the origin. The solution space for an inhomogeneous equation can be expressed as $M \oplus P$, where $P$ denotes a particular solution to the inhomogeneous equation that contains a finite number of degrees of freedom (spin-tensor degrees of freedom of $C_{\al(n-2)}$). Using gauge symmetry transformations, we can identify the general solution with $P$. This justifies the naming of these sets as topological fields.

It may look surprising that the second term in \eqref{Df_decomp} drops out of the equation (since $D^L \omega_{\al(n)} = -{h_\al}^\ad D^L f_{\al(n-1),\ad}$ and ${h_\al}^\ad {h_\al}^\ad F_{\al(n-2),\ad\ad} = 0$). This, however, is analogous to the observation that the equation $\dd A = W$ for a one-form $A$ does not include the symmetric component of the differential, i.e. $\partial_n A_m + \partial_m A_n$.

It should be noted that the non-trivial $\omega_{\al(n)}$ ($n \geq 2$) entails the non-triviality of $\omega_{\al(n+k),\ad(k)}$. This is a consequence of the equation {\eqref{eq:omega}}, which takes the following form in terms of the relevant components ($k \geq 1$):
\begin{equation}
    D^L \omega_{\al(n+k),\ad(k)} - ik(n+k) h_{\al\ad}\omega_{\al(n+k-1),\ad(k-1)} + ih^{\al\ad}\omega_{\al(n+k+1),\ad(k+1)} = 0.
\label{Xeqn45-4.3}
\end{equation}

Another question is whether there exists a field redefinition that eliminates the \rhs\, of \eqref{eq:omega}. If this would be the case, the zero-forms $f_{\al(n-1), \ad}$ were expressed in terms of components of $C_{\al(l), \ad(m)}$. Moreover, these components should then belong to the same module as $C_{\alpha(n-2)}$ on the \rhs\, of \eqref{eq:omega}, being related by the equation {\eqref{eq:C}}. According to \eqref{eq:C}, $l+m = n-2$ in this module. One can make sure that $f_{\al(n-1), \ad}$ cannot be constructed from $C_{\al(n-2-k), \ad(k)}$ ($0\leq k \leq n-2$). Hence, there is no field redefinition that eliminates the \rhs\, of \eqref{eq:omega}.

To complete our analysis, consider equations for $C$ \eqref{eq:C}. They split into sets of independent equations on $C_{\al(l), \ad(m)}$ which belong to modules corresponding to each value of $n = l+m$ hence carrying a finite number of degrees of freedom each, as $l$ and $m$ are bounded for each module. Let us demonstrate this by explicit construction of solutions to the equations.

The $AdS_4$-connection \eqref{AdS-connection} can be represented as follows \cite{Bolotin:1999fa}:
\begin{equation}
    \omega_{AdS} = g^{-1}(y,\y|x)*\dd_x g(y,\y|x).
    \label{gdg}
\end{equation}
A general solution to \eqref{eq:C} then has the form \cite{Didenko:2003aa}
\begin{equation}
    C(y,\y|x) = g^{-1}(y,\y|x) * C_0(y,\y)*g(y,\y|x),
    \label{C:gener_sol}
\end{equation}
where $C_0(y,\y)$ parameterizes initial data.

Consider Poincare coordinates of $AdS_4$ (being of most interest from the holographic perspective \cite{Maldacena:1997re,Gubser:1998bc,Witten:1998qj,Klebanov:2002ja,Giombi:2009wh,Giombi:2010vg,Giombi:2016ejx}):
\begin{equation}
    (\bm{x}^{\al\be}, z), \quad \bm{x}^{\al\be} = \bm{x}^{\be\al},
\label{Xeqn48-4.6}
\end{equation}
where $\bm{x}^{\al\be} = \bm{x}^{\be\al}$ are three-dimensional coordinates of leaves and $z$ is a foliation parameter (the boundary of $AdS_4$ corresponds to the limit $z \rightarrow 0$).

In these coordinates, the $AdS_4$-connection has the following form \cite{Vasiliev:2012vf} (see also \cite{Giombi:2009wh,Giombi:2010vg,Giombi:2016ejx}):
\begin{equation}
    \omega_{AdS} = -\frac{1}{z}\dd \bm{x}^{\al\be} y^-_\al y^-_\be  + \frac{\dd z}{2z} y^+_\al y^{-\al},
\label{Xeqn49-4.7}
\end{equation}
where
\begin{equation}
    y^+_\al = \frac{1}{2}(y_\al - i\y_\al), \quad y^-_\al = \frac{1}{2}(\y_\al - i y_\al)
\label{Xeqn50-4.8}
\end{equation}
with the following properties
\begin{equation}
    f * y^+_\al = \left( y^+_\al - \frac{1}{2} \frac{\partial}{\partial y^{-\al}} \right) f, \quad f * y^-_\al = \left( y^-_\al - \frac{1}{2} \frac{\partial}{\partial y^{+\al}} \right) f,
\label{Xeqn51-4.9}
\end{equation}
\begin{equation}
    y^+_\al * f = \left( y^+_\al + \frac{1}{2} \frac{\partial}{\partial y^{-\al}} \right) f, \quad y^-_\al * f = \left( y^-_\al + \frac{1}{2} \frac{\partial}{\partial y^{+\al}} \right) f.
\label{Xeqn52-4.10}
\end{equation}

To find the representation \eqref{gdg}, it is convenient to consider the equation $\dd_x g = g * \omega_{AdS}$. It admits the following solution \cite{Giombi:2010vg,Didenko:2012tv,Iazeolla:2020jee}:
\begin{equation}
    g = \frac{4\sqrt{\rho z}}{(1+\sqrt{\rho z})^2} \exp \left[ -\frac{4\rho}{(1+\sqrt{\rho z})^2}\, \bm{x}^{\al\be} y^-_\al y^-_\be - 2 \frac{1-\sqrt{\rho z}}{1+\sqrt{\rho z}}\, y^+_\al y^-_\be \epsilon^{\al\be} \right]
    \label{g_Poincare}
\end{equation}
with the parameter $0 < \rho < \infty$ and the inverse
\begin{align}
    g^{-1} = \frac{4\sqrt{\rho z}}{(1+\sqrt{\rho z})^2} \exp \left[ \frac{4\rho}{(1+\sqrt{\rho z})^2}\, \bm{x}^{\al\be} y^-_\al y^-_\be + 2 \frac{1-\sqrt{\rho z}}{1+\sqrt{\rho z}}\, y^+_\al y^-_\be \epsilon^{\al\be} \right].
    \label{g_inv}
\end{align}
The dimensionful parameter can be identified with the curvature of the $AdS$ space upon appropriate rescaling of the frame field, that has to be dimensionless.

In these coordinates, \eqref{C:gener_sol} acquires the following form (this formula can be obtained by the substitution of $y^\pm \rightarrow g^{-1}*y^\pm*g$ in the arguments of $C_0$, as well as through direct computation with some $C_0$ according to \eqref{C:gener_sol}):
\begin{equation}
    C(y^+, y^-|\bm{x}, z) = C_0(\sqrt{\rho z}\, y^+_\al + \frac{2\rho}{\sqrt{\rho z}}\, {\bm{x}_\al}^\be y^-_\be, \frac{1}{\sqrt{\rho z}}\, y^-_\al).
    \label{C:general_sol}
\end{equation}
We see that each independent module carries a finite number of degrees of freedom that parametrized by constants ${C_0}_{\,\al(n-k), \ad(k)}$ with fixed $n$ and $0 \leq k \leq n$.

It is important to note that \eqref{eq:C} represents equations for so-called HS Killing tensors (see e.g. \cite{Bekaert:2005ka}), that describe global symmetries of free fields in  $AdS$ space.  Indeed, the gauge transformation for $\omega_{AdS}$, subject to \eqref{flatness}, has the form $\delta \omega_{AdS} = \dd_x \epsilon + \omega_{AdS}*\epsilon - \epsilon*\omega_{AdS}$ with an arbitrary $K$-independent $\epsilon$.  This observation, in particular, leads to a scheme for the construction of charges in the HS theory using the topological sector, as described in \cite{Didenko:2015pjo}.

\section{Action}
\label{sec5}

An action for free HS fields can be constructed in terms of the curvature \cite{Vasiliev:1986zej}. Remarkably, such a construction is lifted up to the third-order action for interacting HS fields \cite{Fradkin:1986qy}. In this section, we consider an action for free topological fields (that was originally suggested in \cite{Vasiliev:1987hv}, see also \cite{Vasiliev:1988mf}) and construct a cubic action for interacting HS and topological fields.

To construct an action, let us consider the problem of finding an invariant functional in a four-dimensional space. We need to find a four-form to be integrated. There is a curvature two-form
\begin{equation}
    R = \dd_x \omega + \omega*\omega
    \label{Curvature}
\end{equation}
that is suitable for this purpose. It splits into two parts:
\begin{equation}
    R = R^{odd} + R^{even},
\label{Xeqn57-5.2}
\end{equation}
where $R^{odd}$ is odd in $K$ and $R^{even}$ is even in $K$.

By using the supertrace operation \cite{Vasiliev:1986qx}:
\begin{equation}
    str(f(Y,K)) = f(0, 0),
    \label{trace}
\end{equation}
we can consider the following invariant functional:
\begin{equation}
    S^{top} = \frac{1}{2} \int str(R*R).
    \label{Topological_action}
\end{equation}
It is topological, as the variation of \eqref{Topological_action} is identically zero:
\begin{align}
    \delta S^{top}& = \int str((\dd_x\delta\omega+\omega*\delta\omega+\delta\omega*\omega)*R) = \int str(\delta\omega*(\dd_x R + \omega*R-R*\omega)) = 0,
\end{align}
where we used the following property of the supertrace:
\begin{equation}
    str(f(Y)*g(Y)) = str(g(-Y)*f(Y)) = str(g(Y)*f(-Y)),
\label{Xeqn60-5.6}
\end{equation}
and took into account that $f_{\al(n),\ad(m)}$ is a Grassmann even (odd) element for even (odd) values of $n+m$ (due to the spin of the fields, which is equal to $\frac{1}{2}(n+m)+1$ that is a direct consequence of the definition given above {\eqref{omega_expansion}} and {\eqref{TheFirstOn-shell:omega}}).

To construct the action with a nontrivial variation leading to equations of motion, we should modify the structure of \eqref{Topological_action}. Let us consider an action of the form
\begin{equation}
    S = \frac{1}{2} \int str(R*\tilde{R}),
    \label{ac:full_cubic}
\end{equation}
where we use the tilde notation, which was introduced in \cite{Tatarenko:2024csa} for the physical HS fields,
\begin{equation}
    \tilde{R}^{odd} = \sum_{n,m} \frac{a(n, m)}{n!m!} R_{\al(n), \ad(m)}^{k} y^{\al_1}\dots y^{\al_n} \y^{\ad_1}\dots \y^{\ad_m}k + h.c.,
    \label{R_odd_tilde}
\end{equation}
\begin{equation}
    \tilde{R}^{even} = \sum_{n,m} \frac{b(n, m)}{n!m!} y^{\al_1}\dots y^{\al_n} \y^{\ad_1}\dots \y^{\ad_m}(R_{\al(n), \ad(m)} + R_{\al(n), \ad(m)}^{k\bar{k}} k\bar{k}),
\label{Xeqn63-5.9}
\end{equation}
and $R_{\al(n), \ad(m)}$, $R_{\al(n), \ad(m)}^{k}$ and $R_{\al(n), \ad(m)}^{k\bar{k}}$ are the relevant components of $R$ ($R_{\al(n), \ad(m)}^{\bar{k}}$ are conjugate to  $R_{\al(m), \ad(n)}^{k}$ and we omit them in our formulas, using the notation "$h.c.$").

In the sector of fields $\omega$ that are odd in $K$, the quadratic part of \eqref{ac:full_cubic} is
\begin{equation}
    S = \frac{1}{2} \sum_{n,m}\frac{a(n, m)}{n!m!}\int R^{k(1)}_{\al(n),\ad(m)}R^{k(1)\al(n),\ad(m)} + h.c.,
    \label{Action:arbitrary}
\end{equation}
where $R^{k(1)}_{\al(n),\ad(m)}$ is the linearized part of $R^{k}_{\al(n),\ad(m)}$:
\begin{equation}
    R^{(1)} = D_{AdS} \omega,\quad D_{AdS} F :=  d_x F + [\omega_{AdS}, F]_{\pm *},
\label{Xeqn65-5.11}
\end{equation}
where $[a, b]_{\pm *} = a * b \pm b * a$ is a graded (with respect to degrees of differential forms $a, b$) commutator.

Variation of \eqref{Action:arbitrary} entails
\begin{multline}
    \delta S = \sum_{n,m}\frac{i}{n!m!} \int \quad \delta \omega^{\al(n),\ad(m)}\big( nm \big[a(n,m) - a(n-1,m-1)\big] h_{\al\ad} R^{k(1)}_{\al(n-1),\ad(m-1)} +\\+ \big[a(n+1,m+1) - a(n,m)\big] h^{\al\ad} R^{k(1)}_{\al(n+1),\ad(m+1)}  \big) + h.c.\, .
\end{multline}
Let
\begin{equation}
    a(n, m) = \delta(n m),
    \label{a(n,m)}
\end{equation}
where $\delta(n)$ is a Kronecker delta ($\delta(n) = 1$ for $n = 0$; $\delta(n) = 0$ for $n \neq 0$).
Then the action \eqref{Action:arbitrary} is a functional of $\omega^{\al(n)}, \omega^{\al(n),\ad}; \omega^{\ad(m)}, \omega^{\al,\ad(m)}$, and equations of motion acquire the following form:
\begin{equation}
    h^{\al\ad} R^{k(1)}_{\al(n+1),\ad} = 0, \quad \quad h_{\al\ad} R^{k(1)}_{\al(n)} = 0;
\label{Xeqn67-5.14}
\end{equation}
\begin{equation}
    h^{\al\ad} R^{k(1)}_{\al,\ad(m+1)} = 0, \quad \quad h_{\al\ad} R^{k(1)}_{\ad(m)} = 0,
\label{Xeqn68-5.15}
\end{equation}
and $h.c.$.

Note that these equations contained in \eqref{eq:omega} restrict $R^{k(1)}_{\al(n)}$ (and $R^{k(1)}_{\ad(m)}$) to zero modulo terms proportional to $H_{\al\al}C_{\al(n-2)}$ (and $\bar{H}_{\ad\ad}C_{\ad(n-2)}$).

Now consider the full set of fields of the HS theory. In addition to the $K$-odd topological fields $\omega$, they also include $K$-even physical HS fields $\omega$. It was shown in \cite{Fradkin:1986qy} that $b(n,m) = sgn(n-m)$ (with the convention $sgn(0) = 0$) leads to a gauge invariant cubic action for HS fields. Let us demonstrate that this choice, along with \eqref{a(n,m)}, leads to a cubic action of the form \eqref{ac:full_cubic} for interacting HS physical and topological fields. The question to be addressed is whether the action \eqref{ac:full_cubic} is gauge invariant. More precisely, one should prove the existence of some gauge transformation of the form
\begin{equation}
    \delta \omega = D_{AdS} \epsilon + [\omega, \epsilon]_* + \Delta(\epsilon, \omega),
    \label{deformed_gauge}
\end{equation}
that leaves the action \eqref{ac:full_cubic} invariant with some $\Delta(\epsilon, \omega)$ linear in $\omega$.

A gauge variation of the cubic action is of the second order in fields. Using the fact that $\delta R = [R, \epsilon]_*$ under \eqref{deformed_gauge} with $\Delta = 0$, we conclude that
\begin{equation}
    \delta S = \int str([R, \epsilon]_**\tilde{R}) + \frac{\delta S^{(2)}}{\delta \omega} \Delta(\epsilon, \omega) + \text{cubic terms},
    \label{action:vaiation}
\end{equation}
where $S^{(2)}$ is the quadratic part of the action. On shell, the first term consists of products of non-zero parts of the curvatures that are proportional to $H_{\al\be }C$ or $\bar{H}_{\ad\bd}C$. In fact, $H_{\al\be}\bar{H}_{\ad\bd} = 0$, because it is a four-form in a four-dimensional space that is unique up to a numerical factor and does not carry any indices. Such a non-zero form is $H_{\al\be} H^{\al\be} = -\bar{H}_{\ad\bd}\bar{H}^{\ad\bd}$. Thus, we are left with the set of terms of the structure $H_{\al\be }C \times H_{\g\delta}C$ and $\bar{H}_{\ad\bd }C \times \bar{H}_{\gd\deld}C$. To analyze their contribution, split the gauge transformation into even and odd in $K$ parts: $\epsilon = \epsilon^{(even)} + \epsilon^{(odd)}$. Then the gauge transformation of the action with respect to the even part consists of the following terms:
\begin{multline}
    \int str([R^{(even)}, \epsilon^{(even)}]_**\tilde{R}^{(even)}) = \sum_{n}\frac{sgn(n - 0)}{n!}\left(\frac{i\bar{\eta}}{2} \right)^2 \times \\ \times \int ({[H_{\be\be}\partial^{\be}\partial^{\be}C^{(odd)}(y,0), \epsilon^{(even)}]_*)}_{\al(n)} H_{\al\al}C^{(odd)\al(n+2)}\,|_{K=0} + h.c.
    \label{gauge_variate_cubic_action_1}
\end{multline}
and
\begin{multline}
    \int str([R^{(odd)}, \epsilon^{(even)}]_**\tilde{R}^{(odd)}) = \sum_{n}\frac{a(n, 0)}{n!}\left(\frac{i\bar{\eta}}{2} \right)^2 \times \\ \times \int ({[H_{\be\be}y^{\be}y^{\be}C^{(even)}(y,0), \epsilon^{(even)}]_*)}_{\al(n)} H^{\al\al}C^{(even)\al(n-2)}\,|_{K=0} + h.c.\, .
    \label{gauge_variate_cubic_action_2}
\end{multline}
The gauge transformation with respect to the odd part consists of
\begin{multline}
    \int str([R^{(even)}, \epsilon^{(odd)}]_**\tilde{R}^{(odd)})  =\sum_{n}\frac{a(n, 0)}{n!}\left(\frac{i\bar{\eta}}{2} \right)^2 \times \\ \times \int ({[H_{\be\be}\partial^{\be}\partial^{\be}C^{(odd)}(y,0), \epsilon^{(odd)}]_*)}_{\al(n)} H^{\al\al}C^{(even)\al(n-2)}\,|_{K=0} + h.c.
    \label{gauge_variate_cubic_action_3}
\end{multline}
and
\begin{multline}
    \int str([R^{(odd)}, \epsilon^{(odd)}]_**\tilde{R}^{(even)}) = \sum_{n}\frac{sgn(n - 0)}{n!}\left(\frac{i\bar{\eta}}{2} \right)^2 \times \\ \times \int ({[H_{\be\be}y^{\be}y^{\be}C^{(even)}(y,0), \epsilon^{(odd)}]_*)}_{\al(n)} H_{\al\al}C^{(odd)\al(n+2)}\,|_{K=0}  + h.c.\, .
    \label{gauge_variate_cubic_action_4}
\end{multline}
There are independent of $\bar{y}$ (or $y$) parts of the curvature to be commuted with $\epsilon$ in \eqref{gauge_variate_cubic_action_1}--\eqref{gauge_variate_cubic_action_4}. That is why only $\epsilon$ independent of $\bar{y}$ (or $y$) contributes to \eqref{gauge_variate_cubic_action_1}--\eqref{gauge_variate_cubic_action_4}. Due to the constancy (as functions of $n$) of $sgn(n - 0)$ and $a(n, 0)$, \eqref{gauge_variate_cubic_action_1}--\eqref{gauge_variate_cubic_action_4} are absorbed into a gauge variation of the topological action \eqref{Topological_action},
that is equal to\break zero:
\begin{equation}
    \delta S^{top} = \int str([R,\epsilon]_**R) = \int str(R*\epsilon*R - \epsilon*R*R) = 0.
\label{Xeqn71-5.22}
\end{equation}

Therefore, we conclude that the first term in \eqref{action:vaiation} equals zero on the free equations of motion. Consequently,
\begin{equation}
    \delta S = \frac{\delta S^{(2)}}{\delta \omega} G(\omega, \epsilon) + \frac{\delta S^{(2)}}{\delta \omega} \Delta(\epsilon, \omega) + \text{cubic terms}
\label{Xeqn72-5.23}
\end{equation}
with some $G$. Thus, one can set $\Delta = - G$ and obtain gauge invariance of the action \eqref{ac:full_cubic}.

Following \cite{Tatarenko:2024csa}, define the conserved current
\begin{equation}
    J_{\xi} = str\big( \xi(Y,K)*[\omega(Y, K), \tilde{R}^{(1)}(Y, K)]_* \big),
    \label{Current}
\end{equation}
where $\xi$ is a zero-form such that
\begin{equation}
    D_{AdS} \xi = 0.
    \label{Dxi}
\end{equation}
To verify that \eqref{Current} is conserved, consider the variation of the action \eqref{ac:full_cubic} under a gauge transformation \eqref{deformed_gauge}, with an arbitrary $\epsilon$, at the second order in fields relevant for the cubic\break action:
\begin{align}
    \delta S &= \int str (D_{AdS}(\delta\omega) * \tilde{R}+ (\omega*\delta\omega + \delta\omega*\omega)*\tilde{R}^{(1)}) = \int str (\delta\omega*(D_{AdS}\tilde{R} + [\omega, \tilde{R}^{(1)}]_*)) =\notag\\&= \int str \big( (D_{AdS} \epsilon + [\omega, \epsilon]_* + \Delta(\epsilon, \omega))*(D_{AdS}\tilde{R} + [\omega, \tilde{R}^{(1)}]_*) \big) = \notag\\& = \int str \big( ([\omega, \epsilon]_* + \Delta(\epsilon, \omega))*(D_{AdS}\tilde{R} + [\omega, \tilde{R}^{(1)}]_*) \big) - \int str ( \epsilon*D_{AdS}[\omega, \tilde{R}^{(1)}]_*).
\end{align}
Here, the first integral equals zero on the equations of motion, as its integrand is proportional to $\frac{\delta S}{\delta \omega}$. Therefore, on shell, from the gauge invariance of the action, it follows that
\begin{equation}
    \int str ( \epsilon*D_{AdS}[\omega, \tilde{R}^{(1)}]_*) = 0,
\label{Xeqn75-5.27}
\end{equation}
whence
\begin{equation}
    D_{AdS} J_{\xi} = 0,
\label{Xeqn76-5.28}
\end{equation}
and this conservation law can be treated as a consequence of the Noether identity (see \cite{Tatarenko:2024csa} for details).

\eqref{Current} allows one to construct the conserved charges of the following form:
\begin{equation}
    Q = \int_{\Sigma^3} J_{\xi},
\label{Xeqn77-5.29}
\end{equation}
where $\Sigma^3$ is any three-dimensional hypersurface in a four-dimensional space-time.

\section{Conclusion}

In this paper, the topological sector of the nonlinear system of equations for HS fields in four dimensions was considered. It determines the dynamics of one-forms $\omega$ and zero-forms $C$, respectively, odd and even in Klein operators. It is shown that in $AdS$ vacuum, this set does not carry field degrees of freedom, justifying the name "topological fields". In particular, such $\omega$ are pure gauge when $C = 0$, becoming nontrivial topological fields at $C \neq 0$. Note that any independent module in the one-form sector contains an infinite tower of topological fields $\omega_{\al(n+k),\ad(k)}$ (or $\omega_{\al(k),\ad(n+k)}$) with a fixed $n \geq 2$ and an arbitrary $k \geq 0$, which are associated with a finite number of zero-forms $C_{\al(n-2-i),\ad(i)}$ with $i \leq n-2$. Also, we constructed the cubic action for interacting HS and topological fields and derived conserved charges associated with its symmetries.

\section*{Acknowledgements}
The author is grateful to M.A. Vasiliev for fruitful and motivating discussions and comments on the draft. Also the author thanks Yu.A. Tatarenko, A.A. Tarusov and K.A. Ushakov for usefull discussions and V.E. Didenko for valuable comments on the manuscript. The work was supported by the Foundation for the Advancement of Theoretical
Physics and Mathematics ``BASIS''.

\appendix

\section{Derivation of linear equations}
    \label{App1}
    An effective tool for solving the system \eqref{base_all} is the differential homotopy approach \cite{Vasiliev:2023yzx}.
    The idea is to solve \eqref{eq_dd} with the following ansatz:
    \begin{equation}
        f_{\mu} = \fint{p r \tau \sigma^l}\mu_0 \mu(\tau, \sigma^l)\dd \Omega^2  \E(\Omega) g_1(r_1, \y, K)\bar{*}\dots \bar{*}g_n(r_n, \y, K)k,
        \label{anz}
    \end{equation}
    where $l \in \overline{1, n}$, $g_l$ are functions $\omega$ or $C$ with implicit dependence on space-time coordinates $x$;
    $\{t^a\} = \{\tau, \sigma^l\}$ are integration parameters restricted to a compact domain by the measure factor of $\mu(\tau, \sigma^l)$,
    \begin{equation}
            \Omega^{\alpha} = \tau z^{\alpha}-(1-\tau)p^{\alpha}(\sigma), \quad \dd \Omega^2 = \dd \Omega^{\alpha} \dd \Omega_{\alpha}, \quad \dd = \theta^\al \frac{\partial}{\partial z^\al} + \dd t^a \frac{\partial}{\partial t^a},
    \end{equation}
    \begin{equation}
        \begin{gathered}
            {\E}({\Omega})  = \exp i \left({\Omega}_{\alpha}(y+p_+)^{\alpha} - \sum_{i<j} p_{i\alpha} p_j^{\alpha} - \sum_{i=1}^n p_{i\alpha} r_i^{\alpha}  \right),
        \end{gathered}
    \end{equation}
    \begin{equation}
        p_+^{\alpha} = \sum_{i=1}^n p_i^{\alpha}, \quad p^{\alpha}(\sigma) = \sum_{i=1}^n p_i^{\alpha} \sigma^i, \quad \mu_0 = \frac{1}{(2\pi)^n}\prod_{i=1}^{n}(\dd^2p_i \dd^2r_i),
    \end{equation}
    $\bar{*}$ denotes the $*$-product \eqref{star} with respect to antiholomorphic spinor variables $\y_{\ad}$ on which $g_l$ depend.

    The main property of \eqref{anz} is
    \begin{equation}
        \dd_Z f_{\mu} = (-1)^{n+1}f_{\dd \mu}.
        \label{d_measure}
    \end{equation}

    With the aid of the ansatz \eqref{anz}, we solve the equation on $S_1$
    \begin{equation}
        -2i\dd_Z S_1 = i\eta C*\gamma + i\bar{\eta} C*\bar{\gamma}
        \label{eq:S_1}
    \end{equation}
     in the following form \cite{Vasiliev:2023yzx}:
    \begin{equation}
        S_1 = -\frac{\eta}{2}\fint{ p r \tau \sigma} \mu_0 l(\tau) \D(\sigma) \dd \Omega^2 \E(\Omega) C(r, \bar{y}, K) k -\frac{\bar{\eta}}{2}\fint{ \bar{p} \bar{r} \bar{\tau} \bar{\sigma}} \mu_0 l(\bar{\tau}) \D(\bar{\sigma}) \dd \bar{\Omega}^2 \E(\bar{\Omega}) C(y, \bar{r}, K) \bar{k},
        \label{S1}
    \end{equation}
    \begin{equation}
        l(\xi) = \theta(\xi)\theta(1-\xi), \quad \D(\xi) = \delta(\xi) \dd \xi .
    \end{equation}
    Then, the equation on $W_1$
    \begin{equation}
        2i\dd_Z W_1 = \dd_x S_1 + \{\omega_{AdS}, S_1\}_*
    \end{equation}
    can be solved as
    \begin{multline}
        W_1= \omega(Y,K) + \dfrac{i\eta}{4}\fint{p r \tau \sigma \sigma^{\omega} } \mu_0 l(\tau) \D(\sigma)  \dd \Omega^2 \E(\Omega) \big\{ P(-1, \sigma^{\omega}, \sigma) \omega_{AdS}(r_{\omega}, \bar{y}, K) \bar{*} C(r, \bar{y}, K) +\\+ P(\sigma, \sigma^{\omega}, 1)  \dd \Omega^2 \E(\Omega) C(r, \bar{y}, K) \bar{*} \omega_{AdS}(r_{\omega}, \bar{y}, K) \big \} k +\\+\dfrac{i\bar{\eta}}{4}\fint{\bar{p} \bar{r} \bar{\tau} \bar{\sigma} \bar{\sigma}^{\omega} } \mu_0 l(\bar{\tau}) \D(\bar{\sigma})  \dd \bar{\Omega}^2 \E(\bar{\Omega}) \big\{ P(-1, \bar{\sigma}^{\omega}, \bar{\sigma}) \omega_{AdS}(y, \bar{r}_{\omega}, K) \underline{*} C(y, \bar{r}, K) +\\+ P(\bar{\sigma}, \bar{\sigma}^{\omega}, 1)  \dd \bar{\Omega}^2 \E(\bar{\Omega}) C(y, \bar{r}, K) \underline{*} \omega_{AdS}(y, \bar{r}_{\omega}, K)  \big \} \bar{k},
        \label{W_1}
    \end{multline}
    where $\omega(Y,K)$ is $Z$-independent,
    \begin{equation}
        P(a_1, \dots, a_n) = \theta(a_n-a_{n-1})\theta(a_{n-1}-a_{n-2})\dots\theta(a_2-a_1),
    \end{equation}
    and $\underline{*}$ denotes the $*$-product with respect to only holomorphic spinor variables $y_{\al}$.

    The linearized equation on one-forms $\omega$ has the form
    \begin{equation}
        \dd_x \omega +\{\omega_{AdS}, \omega\}_*  + \dd_x W_1 +\{\omega_{AdS}, W_1\}_* = 0,
    \end{equation}
    and after substitution of \eqref{W_1} takes the form \cite{Vasiliev:2023yzx}
    \begin{multline}
        \dd_x \omega +\{\omega_{AdS}, \omega\}_* = \dfrac{\eta}{4i}\fint{p r \tau \sigma \sigma^{\omega1} \sigma^{\omega2}} \mu_0 \D(\tau) \D(\sigma)  \dd \Omega^2 \E(\Omega)\big [ P(-1, \sigma^{\omega1}, \sigma^{\omega2}, \sigma, 1) \omega_{AdS}(r_{\omega 1}, \bar{y}, K) \bar{*} \omega_{AdS}(r_{\omega 2}, \bar{y}, K) \bar{*} C(r, \bar{y}, K) + \\+ P(-1, \sigma^{\omega1}, \sigma, \sigma^{\omega2}, 1) \omega_{AdS}(r_{\omega 1}, \bar{y}, K) \bar{*} C(r, \bar{y}, K) \bar{*} \omega_{AdS}(r_{\omega 2}, \bar{y}, K)+ \\+ P(-1, \sigma, \sigma^{\omega1}, \sigma^{\omega2}, 1) C(r, \bar{y}, K) \bar{*} \omega_{AdS}(r_{\omega 1}, \bar{y}, K) \bar{*} \omega_{AdS}(r_{\omega 2}, \bar{y}, K)  \big] k\, +\\+ \dfrac{\bar{\eta}}{4i}\fint{\bar{p} \bar{r} \bar{\tau} \bar{\sigma} \bar{\sigma}^{\omega1}\bar{\sigma}^{\omega2}} \mu_0 \D(\bar{\tau}) \D(\bar{\sigma})  \dd \bar{\Omega}^2 \E(\bar{\Omega})\big [ P(-1, \bar{\sigma}^{\omega1}, \bar{\sigma}^{\omega2}, \bar{\sigma}, 1) \omega_{AdS}(y,\bar{r}_{\omega 1}, K) \underline{*} \omega_{AdS}(y,\bar{r}_{\omega 2}, K) \underline{*} C(y,\bar{r}, K) + \\+ P(-1, \bar{\sigma}^{\omega1}, \bar{\sigma}, \bar{\sigma}^{\omega2}, 1) \omega_{AdS}(y, \bar{r}_{\omega 1}, K) \underline{*} C(y, \bar{r}, K) \underline{*} \omega_{AdS}(y, \bar{r}_{\omega 2}, K)+ \\+ P(-1, \bar{\sigma}, \bar{\sigma}^{\omega1}, \bar{\sigma}^{\omega2}, 1) C(y, \bar{r}, K) \underline{*} \omega_{AdS}(y, \bar{r}_{\omega 1}, K) \underline{*} \omega_{AdS}(y, \bar{r}_{\omega 2}, K)  \big] \bar{k}\,.
        \label{D_omega}
    \end{multline}

    The substitution of $K$-odd $C$ leads to \eqref{TheFirstOn-shell:omega} and $K$-even $C$ leads to \eqref{eq:omega}.

\section{Some calculations}
\label{App2}
    To prove \eqref{equality_for_gauge}, consider the following steps (recall that $a_\al b^\al = -a^\al b_\al$):
    \begin{multline}
        h^{l\al\ad} D^L_l {h^{m\al}}_{\ad} D^L_m \epsilon_{\al(n)} = h^{l\al\ad} D^L_l ({h^{m\al}}_{\ad}) D^L_m \epsilon_{\al(n)} + h^{l\al\ad} {h^{m\al}}_{\ad}  D^L_l D^L_m \epsilon_{\al(n)} = \\= h^{l\al\ad} D^L_l ({h^{m\al}}_{\ad}) D^L_m \epsilon_{\al(n)} + \frac{1}{2} h^{l\al\ad} {h^{m\al}}_{\ad} ( D^L_l D^L_m - D^L_m D^L_l  ) \epsilon_{\al(n)}.
        \label{base_for_app_B}
    \end{multline}
   Since $D^L D^L = -H^{\al\be}y_\al \partial_\be - \bar{H}^{\ad\bd}\y_\ad \dif_\bd$ (which follows from \eqref{flatness}),
    \begin{equation}
        \frac{1}{2} h^{l\al\ad} {h^{m\al}}_{\ad} ( D^L_l D^L_m - D^L_m D^L_l  ) \epsilon_{\al(n)} = -n h^{l\al\ad} {h^{m\al}}_{\ad} h_{l\al\bd}{h_{m\be}}^\bd {\epsilon^\be}_{\al(n-1)} = 0,
    \end{equation}
    where we used $h^{k\al\ad} h_{k\be\bd} = 2\delta^\al_\be \delta^\ad_\bd$. Using this identity again, one finds
    \begin{equation}
        0 = D^L_l(h^{k\al\ad} h_{k\be\bd}) = D^L_l(h^{k\al\ad}) h_{k\be\bd} + h^{k\al\ad} D^L_l(h_{k\be\bd}),
    \end{equation}
    and contracting with $h^{m\be\bd}$, one obtains
    \begin{equation}
        D^L_l(h^{m\al\ad}) = - \frac{1}{2} h^{m\be\bd} h^{k\al\ad} D^L_l(h_{k\be\bd}).
        \label{Dh_inversed}
    \end{equation}

    Substitution of \eqref{Dh_inversed} into \eqref{base_for_app_B} yields
    \begin{multline}
        h^{l\al\ad} D^L_l ({h^{m\al}}_{\ad}) D^L_m \epsilon_{\al(n)}= - \frac{1}{2}  h^{m\be\bd} h^{l\al\ad}  {h^{k\al}}_\ad D^L_l(h_{k\be\bd})   D^L_m \epsilon_{\al(n)} = \\ = - \frac{1}{4} h^{m\be\bd} h^{l\al\ad}  {h^{k\al}}_\ad (D^L_l(h_{k\be\bd}) -  D^L_k(h_{l\be\bd}) )  D^L_m \epsilon_{\al(n)} = 0,
    \end{multline}
since $D^L  h_{\be\bd} = 0$ which follows from \eqref{flatness}.


\begin{thebibliography}{00}

\bibitem{Bekaert:2004qos}
X.~Bekaert, S.~Cnockaert, C.~Iazeolla and M.~A.~Vasiliev,
[arXiv:hep-th/0503128 [hep-th]].

\bibitem{Bekaert:2010hw}
X.~Bekaert, N.~Boulanger and P.~Sundell,
Rev. Mod. Phys. \textbf{84} (2012), 987-1009
[arXiv:1007.0435 [hep-th]].

\bibitem{Ponomarev:2022vjb}
D.~Ponomarev,
Int. J. Theor. Phys. \textbf{62} (2023) no.7, 146
[arXiv:2206.15385 [hep-th]].

\bibitem{Fronsdal:1978rb}
C.~Fronsdal,
Phys. Rev. D \textbf{18} (1978), 3624
PhysRevD.18.3624

\bibitem{Fang:1978wz}
J.~Fang and C.~Fronsdal,
Phys. Rev. D \textbf{18} (1978), 3630


\bibitem{Didenko:2026nag}
V.~E.~Didenko,
[arXiv:2601.10680 [hep-th]].

\bibitem{Sagnotti:2011jdy}
A.~Sagnotti,
J. Phys. A \textbf{46} (2013), 214006
[arXiv:1112.4285 [hep-th]].

\bibitem{Gaberdiel:2015wpo}
M.~R.~Gaberdiel and R.~Gopakumar,
JHEP \textbf{09} (2016), 085
[arXiv:1512.07237 [hep-th]].

\bibitem{Vasiliev:2018zer}
M.~A.~Vasiliev,
JHEP \textbf{08} (2018), 051
[arXiv:1804.06520 [hep-th]].

\bibitem{Tarusov:2025sre}
A.~A.~Tarusov, K.~A.~Ushakov and M.~A.~Vasiliev,
JHEP \textbf{08} (2025), 052
[arXiv:2503.05948 [hep-th]].

\bibitem{Didenko:2025xca}
V.~E.~Didenko and I.~S.~Faliakhov,
Phys. Rev. D \textbf{112} (2025) no.10, 106010
[arXiv:2509.01477 [hep-th]].

\bibitem{Tarusov:2026ich}
A.~A.~Tarusov and K.~A.~Ushakov,
[arXiv:2602.10788 [hep-th]].

\bibitem{Vasiliev:1992av}
  M.~A.~Vasiliev,
  \textit{Phys.~Lett.~B \textbf{285}, 225},
  1992.

\bibitem{Vasiliev:2025erl}
M.~A.~Vasiliev and V.~A.~Vereitin,
[arXiv:2508.11500 [hep-th]].

\bibitem{Didenko:2015pjo}
V.~E.~Didenko, N.~G.~Misuna and M.~A.~Vasiliev,
JHEP \textbf{03} (2017), 164
JHEP03(2017)164
[arXiv:1512.07626 [hep-th]].

\bibitem{Vasiliev:1987hv}
M.~A.~Vasiliev,
Nucl. Phys. B \textbf{307} (1988), 319


\bibitem{Vasiliev:1988sa}
M.~A.~Vasiliev,
Annals Phys. \textbf{190} (1989), 59-106


\bibitem{Bolotin:1999fa}
K.~I.~Bolotin and M.~A.~Vasiliev,
Phys. Lett. B \textbf{479} (2000), 421-428
[arXiv:hep-th/0001031 [hep-th]].

\bibitem{Didenko:2003aa}
V.~E.~Didenko and M.~A.~Vasiliev,
J. Math. Phys. \textbf{45} (2004), 197-215
[arXiv:hep-th/0301054 [hep-th]].

\bibitem{Maldacena:1997re}
J.~M.~Maldacena,
Adv. Theor. Math. Phys. \textbf{2} (1998), 231-252
[arXiv:hep-th/9711200 [hep-th]].

\bibitem{Gubser:1998bc}
S.~S.~Gubser, I.~R.~Klebanov and A.~M.~Polyakov,
Phys. Lett. B \textbf{428} (1998), 105-114
[arXiv:hep-th/9802109 [hep-th]].

\bibitem{Witten:1998qj}
E.~Witten,
Adv. Theor. Math. Phys. \textbf{2} (1998), 253-291
[arXiv:hep-th/9802150 [hep-th]].

\bibitem{Klebanov:2002ja}
I.~R.~Klebanov and A.~M.~Polyakov,
Phys. Lett. B \textbf{550} (2002), 213-219
[arXiv:hep-th/0210114 [hep-th]].

\bibitem{Giombi:2009wh}
S.~Giombi and X.~Yin,
JHEP \textbf{09} (2010), 115
[arXiv:0912.3462 [hep-th]].

\bibitem{Giombi:2010vg}
S.~Giombi and X.~Yin,
JHEP \textbf{04} (2011), 086
[arXiv:1004.3736 [hep-th]].

\bibitem{Giombi:2016ejx}
S.~Giombi,
[arXiv:1607.02967 [hep-th]].

\bibitem{Vasiliev:2012vf}
M.~A.~Vasiliev,
J. Phys. A \textbf{46} (2013), 214013
[arXiv:1203.5554 [hep-th]].

\bibitem{Didenko:2012tv}
V.~E.~Didenko and E.~D.~Skvortsov,
JHEP \textbf{04} (2013), 158
[arXiv:1210.7963 [hep-th]].

\bibitem{Iazeolla:2020jee}
C.~Iazeolla,
PoS \textbf{CORFU2019} (2020), 181
[arXiv:2004.14903 [hep-th]].

\bibitem{Bekaert:2005ka}
X.~Bekaert and N.~Boulanger,
Nucl. Phys. B \textbf{722} (2005), 225-248
[arXiv:hep-th/0505068 [hep-th]].

\bibitem{Vasiliev:1986zej}
M.~A.~Vasiliev,
Fortsch. Phys. \textbf{35} (1987) no.11, 741-770

\bibitem{Fradkin:1986qy}
E.~S.~Fradkin and M.~A.~Vasiliev,
Nucl. Phys. B \textbf{291} (1987), 141-171

\bibitem{Vasiliev:1988mf}
M.~A.~Vasiliev,
Sov. J. Nucl. Phys. \textbf{47} (1988), 531-537

\bibitem{Vasiliev:1986qx}
M.~A.~Vasiliev,
Fortsch. Phys. \textbf{36} (1988), 33-62
LEBEDEV-86-290.

\bibitem{Tatarenko:2024csa}
Y.~A.~Tatarenko and M.~A.~Vasiliev,
JHEP \textbf{07} (2024), 246
[arXiv:2405.02452 [hep-th]].

\bibitem{Vasiliev:2023yzx}
M.~A.~Vasiliev,
JHEP \textbf{11} (2023), 048
[arXiv:2307.09331 [hep-th]].
  
\end{thebibliography}
\end{document}